\documentclass[a4paper]{jpconf}
\usepackage{graphicx,amsmath,cite}
\usepackage[utf8]{inputenc}
\usepackage[english]{babel}
\graphicspath{{./figs/}}

\begin{document}
\title{Two- and three-pion Lévy femtoscopy with PHENIX}

\author{Máté Csanád (for the PHENIX Collaboration)}

\address{Eötvös Loránd University, Pázmány P. s. 1/a, H-1117 Budapest, Hungary}

\ead{csanad@elte.hu}

\begin{abstract}
The last decades of high energy physics revealed, that in ultra-relativistic ion-ion collisions, a strongly interacting quark
gluon plasma (sQGP) is created. Varying the collision energy allows for the investigation of the phase diagram of QCD matter.
The nature of the quark-hadron transition can be studied via femtoscopy, as the investigation of momentum correlations in
heavy ion reactions reveals the space-time structure of the hadron production of the sQGP. Going beyond the Gaussian
assumption the shape of this source may be described by Lévy distributions. In this paper we report on recent femtoscopic
measurements of PHENIX, utilizing Lévy sources.
\end{abstract}

\section{Introduction}
Pioneering interferometry discoveries were made by R. Hanbury Brown as he investigated correlations of
intensity fluctuations with radio telescopes~\cite{HanburyBrown:1952na} at the Jordell Bank Observatory, where he was
able to measure the angular diameter of two strong radiofrequency sources. Constructing two tabletop
experiments~\cite{HanburyBrown:1956zza,HanburyBrown:1957na}, Hanbury Brown and R. Q. Twiss prove that these
correlations are also observable in the optical spectrum. Subsequently, the angular diameter of Sirius was
measured~\cite{HanburyBrown:1956pf}, utilizing this technique. These results can be regarded~\cite{Townes:1999apj}
as the extension of the work by A. A. Michelson and F. G. Pease~\cite{Michelson:1921apj}, originally suggested by H. L. Fizeau and
Michelson~\cite{Townes:1999apj}. Brown and collaborators, building an optical interferometer~\cite{HanburyBrown:1967mna}
at the Narrabi Observatory, were able to measure the angular diameter of multiple
stars~\cite{HanburyBrown:1964na,HanburyBrown:1967mnb}. Nowadays, this technique can be applied to study
the two dimensional structure of distant astrophysical objects~\cite{Dravins:2010spi}.

These phenomena were interpreted in the language of quantum optics by
R. J. Glauber~\cite{Glauber:1962tt,Glauber:2006zz,Glauber:2006gd},
and observed in high energy physics independently by G. Goldhaber and collaborators in proton-antiproton
collisions~\cite{Goldhaber:1959mj,Goldhaber:1960sf}.  These Hanbury Brown and Twiss (HBT) correlations
have an important role in high energy physics, as they can be utilized to understand the femtometer
scale space-time structure of the particle emitting source. Hence this field is often called femtoscopy~\cite{Lednicky:2001qv}.
The basic working principle of femtoscopy is, that if we define the probability density of particle creation at space-time point  $x$
and momentum $p$ as $S(r,p)$ (a.k.a. the source), then the momentum correlation function, $C(q,K)$ (where $q$ is the pair momentum difference
and $K$ is the average momentum of the pair)
can be measured and (assuming thermal emission, lack of final state interactions, etc., see e.g.
Ref.~\cite{Adare:2017vig} for details) is related to the source as
\begin{equation}\label{e:Cp1p2:general}
C_2(q,K)= 1+ \left|\frac{\widetilde S(q,K)}{\widetilde S(0,K)}\right|^2,
\end{equation}
where $\widetilde S(q,K)=\int S(x,K) e^{iqx} d^4x$ is the Fourier transformed of the source. It is worth noting that
if we introduce the spatial correlation
function as
\begin{align}
D(x,K) = \int S(r+x/2,K) S(r-x/2,K) d^4r.
\end{align}
then the correlation function can be expressed as
\begin{align}
C_2(q,K)=1+\frac{\widetilde D(q,K)}{\widetilde D(0,K)},
\end{align}
where again $\widetilde D(q,K) = \int D(x,K) e^{iqx} d^4x$ is the Fourier transformed of the spatial correlation function.
Note that in this case, no absolute value is taken, hence no sign or phase information is lost.
Hence, momentum correlations measure the source, or more directly, the spatial correlation function.

The geometry of the source was often assumed to be Gaussian. However, the expansion of the
medium and the corresponding increase of the mean free path suggests the onset of anomalous diffusion~\cite{Metzler:1999zz,Csanad:2007fr}.
This leads to Lévy distributions, which in turn result in stretched exponential type of momentum correlation functions:
\begin{align}
C_2(q,K)=1 + \exp\left(-|q R(K)|^{\alpha(K)}\right),
\end{align}
where $q$ is the momentum difference, $R$ is the $K$ dependent source size (or its length of
homogeneity), and  $\alpha$ is the Lévy-stability exponent, resulting from the anomalous diffusion, and probably
also depending on $K$. The $\alpha$ parameter is interesting also because it may be connected to the critical exponent $\eta$, appearing in the power-law
exponent of the spatial correlation function at the critical point~\cite{Csorgo:2003uv}.
Recently, Lévy sources have been utilized~\cite{Adare:2017vig,Csanad:2017usp,Kincses:2017zlb,Lokos:2018dqq,Novak:2018hqd}
to describe the particle emitting source. In present paper we report on the recent Lévy femtoscopy results of the PHENIX experiment.

Before we move on to the data analysis and the results, let us mention one important point. The above
equations imply, that at zero relative momentum, the correlation function reaches a value of 2, i.e. $C_2(0,K)=2$.
Experimentally, the low-$q$ region is usually inaccessible due to two-track resolution limitations. Hence one can
only extrapolate to measure $C_2(0,K)$, the zero momentum intercept of the two-particle correlation function.
It turns out, that this value is not 2, but has to be described by 
\begin{align}
\lim_{q\rightarrow 0} C_2(q,K) = 1 + \lambda(K),
\end{align}
where $\lambda$ is the ($K$ dependent) intercept parameter or correlation function strength. This observation
can be easily understood in terms of the core-halo picture~\cite{Bolz:1992hc,Csorgo:1994in},
where the source has two components: a hydrodynamically behaving fireball-type core, surrounded
by a halo of the decay products of long lived resonances. In the core-halo picture,
it turns out (as detailed e.g. in Ref.~\cite{Adare:2017vig}) that
\begin{align}
\lambda = f_c^2,\quad f_c = \frac{N_{\rm core}}{N_{\rm core}+N_{\rm halo}},
\end{align}
where $N_{\dots}$ denotes the number of pions produced in a given component of the source.
Hence $\lambda$ carries indirect information on the number of long-lived resonances
decaying to pions, with an important example being the $\eta'$ meson, as discussed in detail
in Refs.~\cite{Kapusta:1995ww,Vance:1998wd,Csorgo:2009pa,Adare:2017vig}.

\begin{figure}
\begin{minipage}{0.495\linewidth}
\begin{center}
\includegraphics[width=1\linewidth]{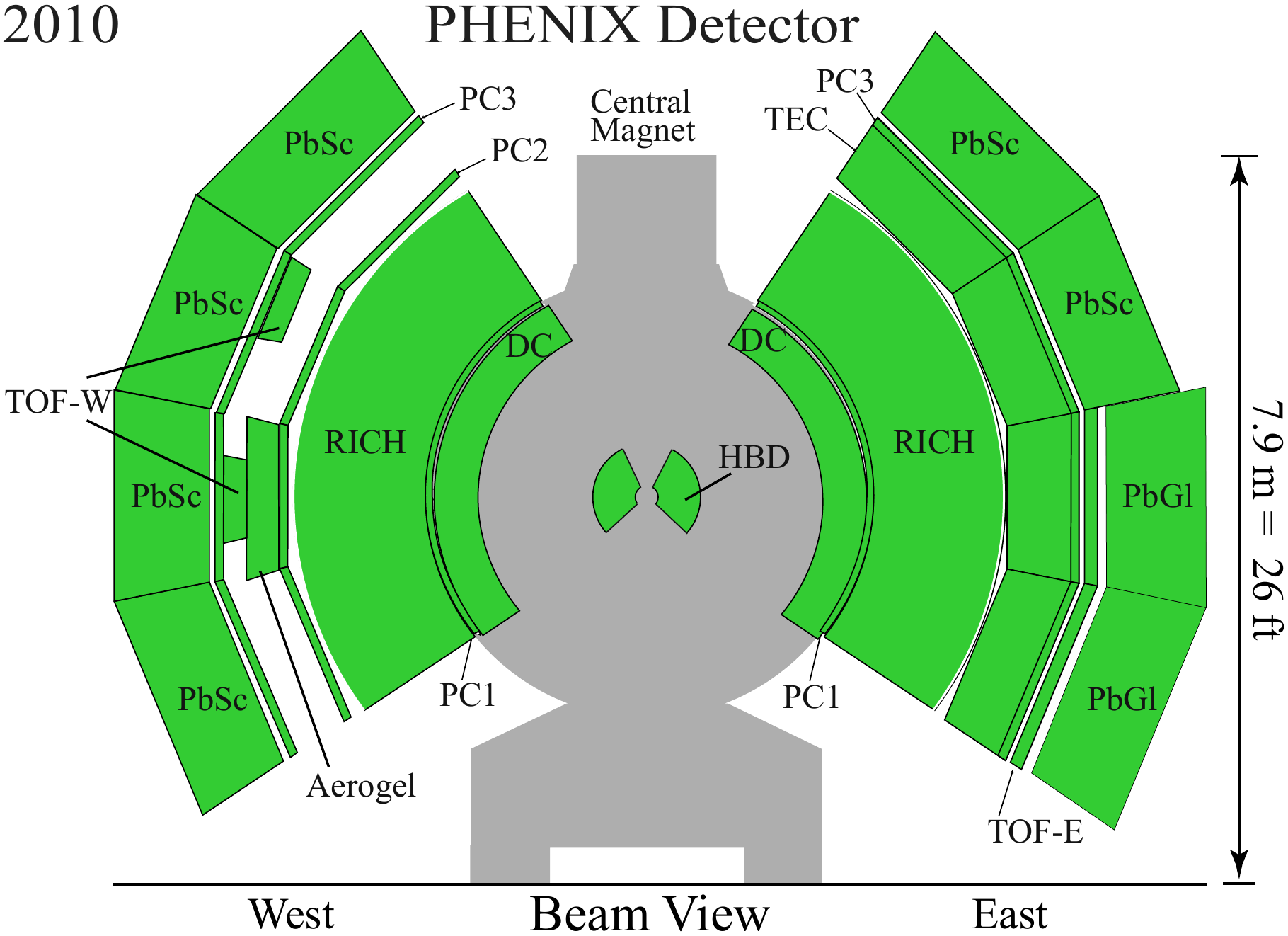}
\end{center}
\caption{The 2010 PHENIX detector setup.}
\label{f:phenix}
\end{minipage}
\begin{minipage}{0.495\linewidth}
\begin{center}
\includegraphics[clip, trim=0cm 0cm 1.1cm 0.93cm, width=1\linewidth]{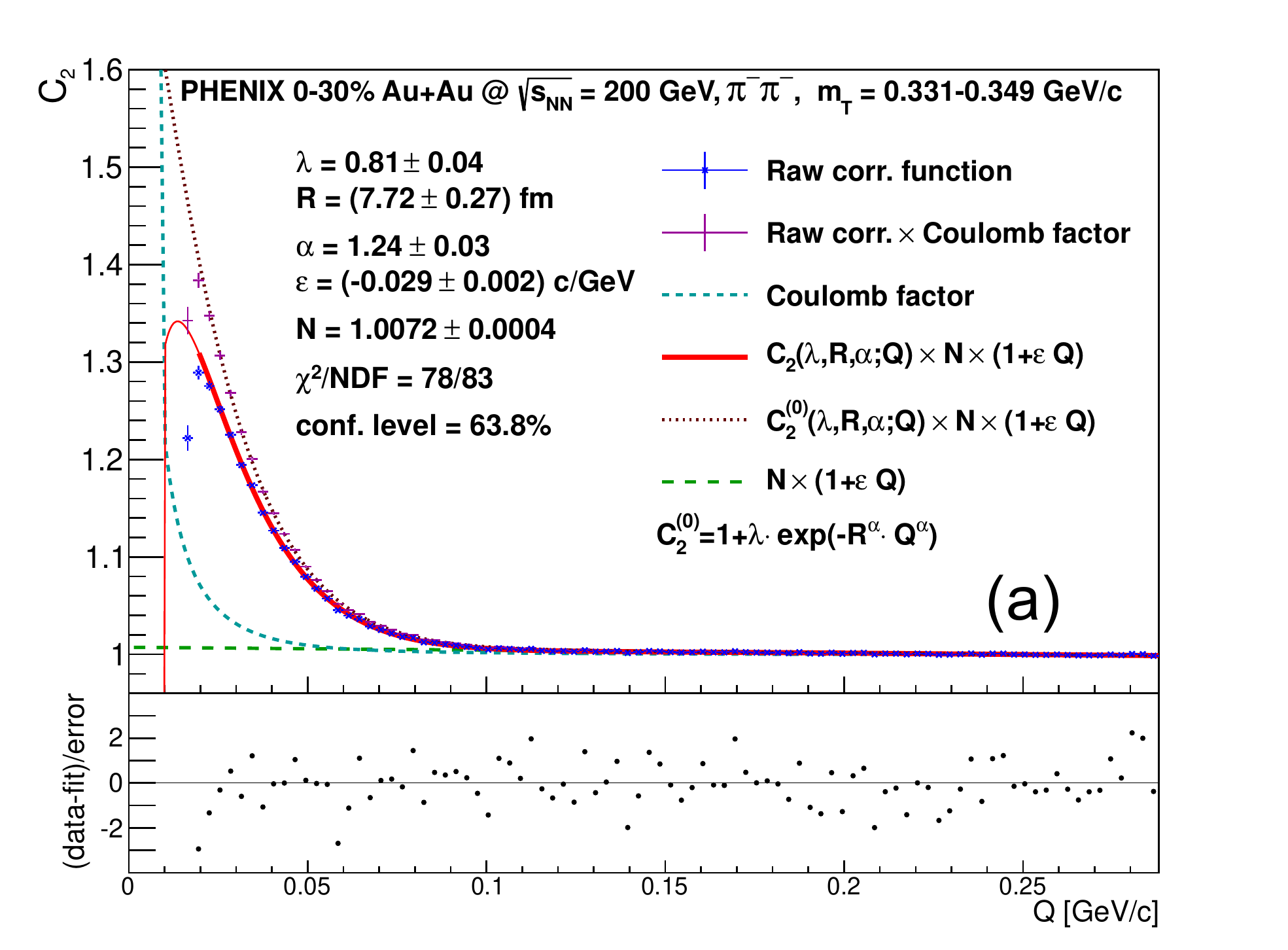}
\end{center}
\caption{Example correlation function.}
\label{f:fit}
\end{minipage}
\end{figure}

\section{Two-pion femtoscopy in 200 GeV Au+Au collisions}
The PHENIX experiment~\cite{Adcox:2004mh} recorded collisions of various nuclei, ranging from protons and deuterons to
gold and uranium, at center of mass per nucleon energies ($\sqrt{s_{_{NN}}}$) ranging from 7.7 to 200 GeV. This allows
to study the QCD phase diagram. In this paper, we discuss the analysis of 0--30\% $\sqrt{s_{_{NN}}}=200$ GeV Au+Au collisions
recorded in the 2010 data taking period. The setup of the experiment during this period is shown in Fig.~\ref{f:phenix}.
In our correlation measurements we applied particle identification via time of flight data from the PbSc and TOF detectors.
In addition to track matching cuts, we applied pair cuts to reduce the effect of track merging and splitting, as detailed
in Refs.~\cite{Adare:2017vig,Csanad:2017usp,Kincses:2017zlb,Lokos:2018dqq,Novak:2018hqd}.

We measured pair momentum difference distributions in the longitudinally comoving system (LCMS) of each pair,
for particles belonging to the same event, as a function of $Q$, the magnitude of the LCMS momentum difference
three-vector. These distributions are usually called actual pair distributions, and are denoted by $A_2(Q,K)$.
To investigate only effects which arise from the pair coming from the same event, we created the same number of
mixed events (with the same multiplicity distribution). Taking a given actual event, we constructed a similar mixed events
by selecting the same number of particles from a large pool of similar background events, making sure all particles
are from different events. This resulted in the so-called background pair distribution, $B_2(Q,K)$. The correlation function
was then calculated as
\begin{align}
C_2(Q,K)=\frac{A_2(Q,K)}{B_2(Q,K)},
\end{align}
separately for positive and negative charges. The $q$-dependence
was investigated in 31 bins of pair transverse mass, related to average pair momentum $K$ as
\begin{align}
m_T = \sqrt{m^2+K_T^2},
\end{align}
where $K_T$ is the pair average transverse momentum. The resulting 62 correlation
functions were fitted assuming Lévy sources, as well as taking into account the Coulomb interaction,
as detailed in Ref.~\cite{Adare:2017vig}. An example fit is shown in Fig.~\ref{f:fit}. We then investigated the main fit parameters.

\begin{figure}
\begin{center}
\includegraphics[width=0.495\linewidth]{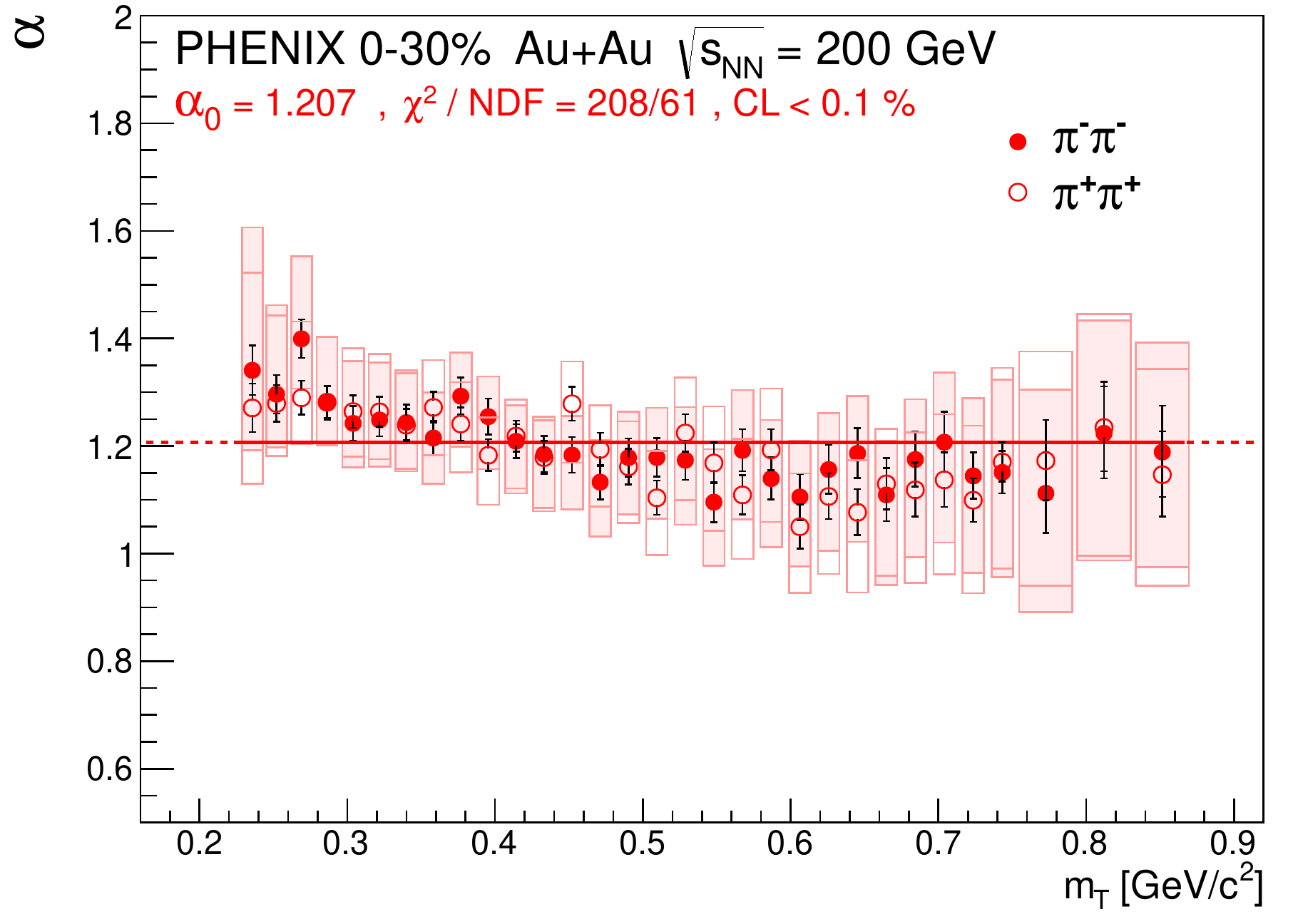}
\includegraphics[width=0.495\linewidth]{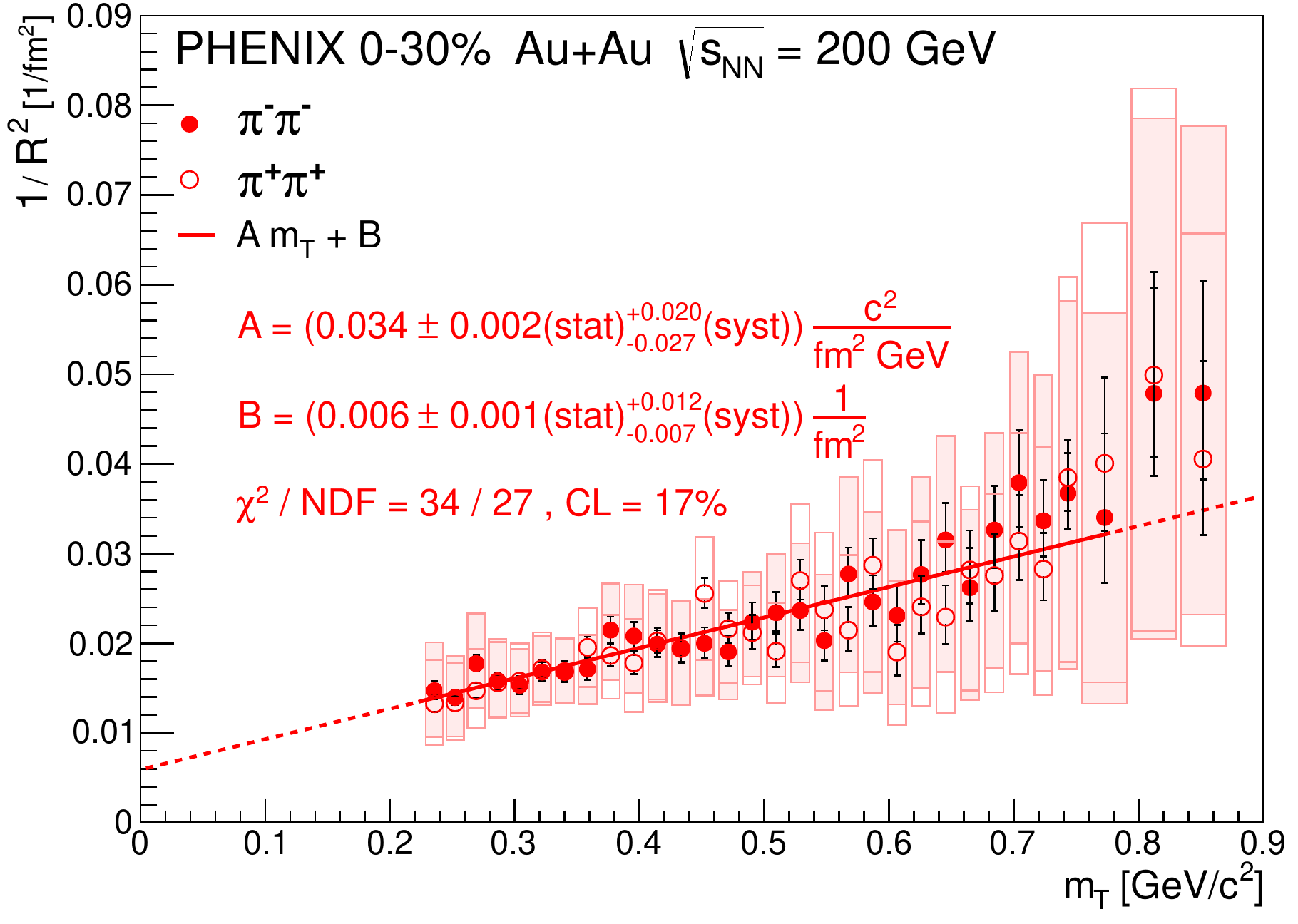}
\includegraphics[width=0.495\linewidth]{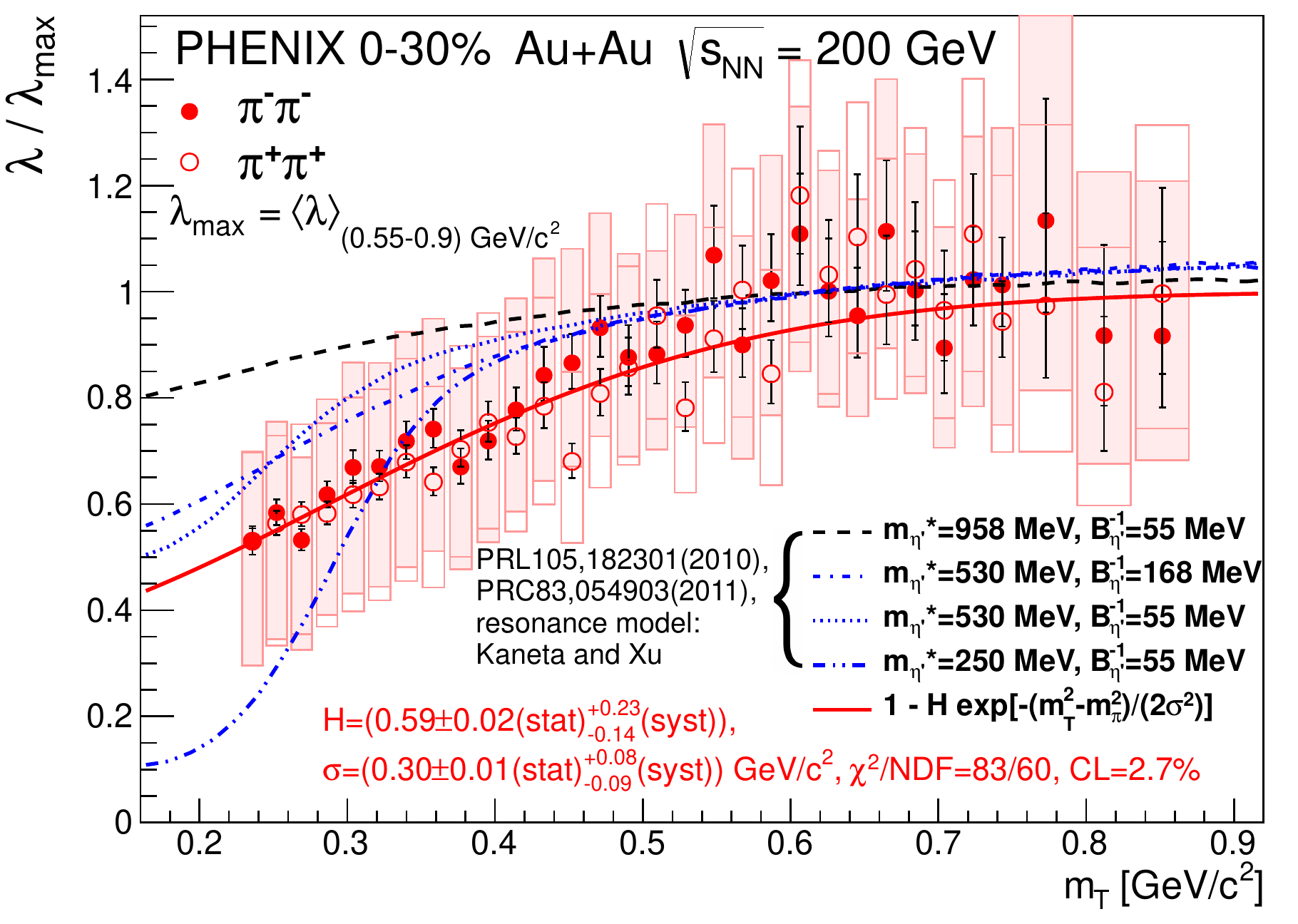}
\includegraphics[width=0.495\linewidth]{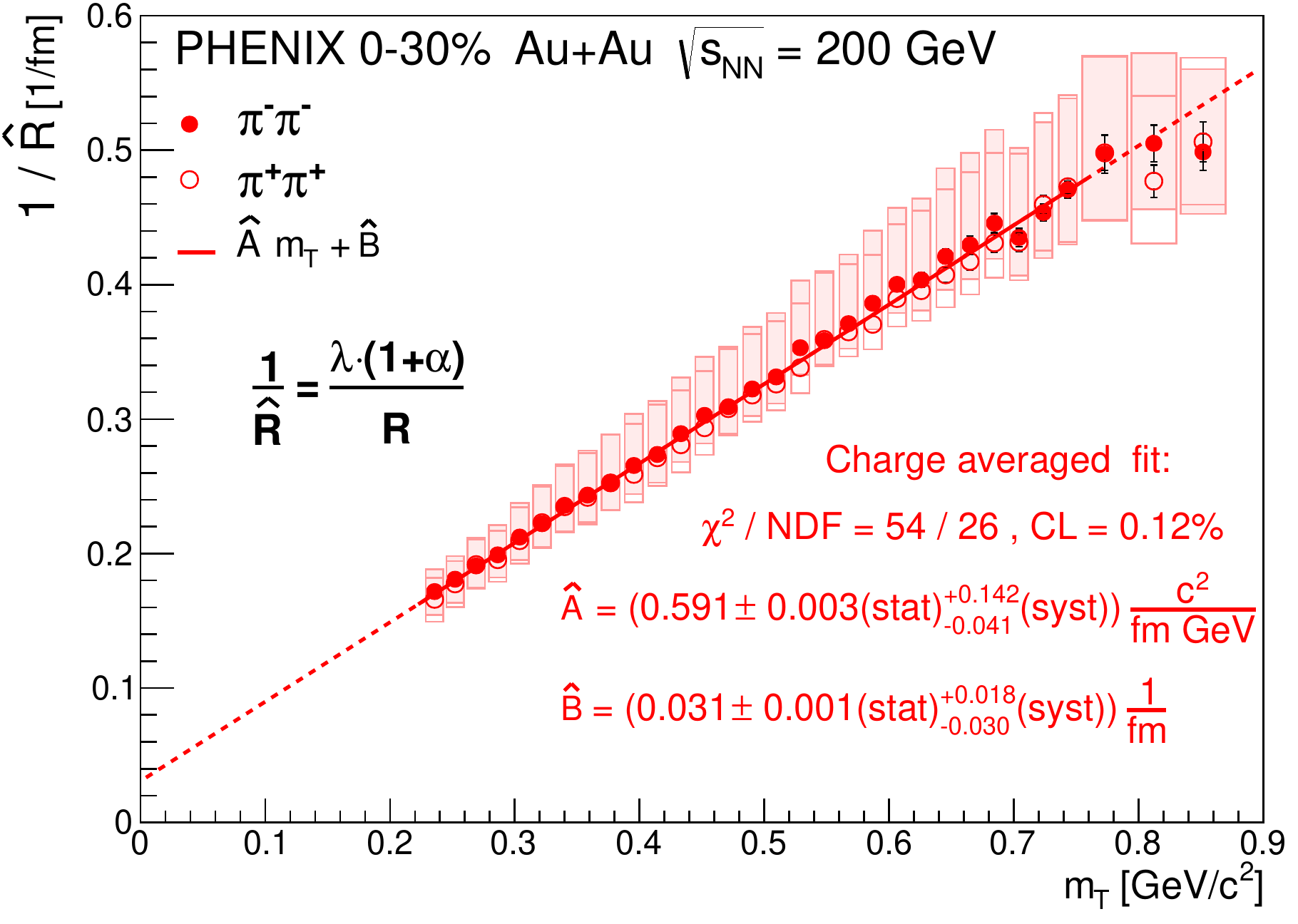}
\end{center}
\caption{L\'evy parameters $\alpha$, $R$ (shown as $1/R^2$) and $\lambda$ (shown normalized as $\lambda/\lambda_{\rm max}$),
as well as the new scaling parameter $\widehat{R}$ versus pair $m_T$ show in the top left, top right, bottom left and bottom right
panels, respectively. Statistical uncertainties are shown as bars, systematic ones with boxes.}
\label{f:0030pars}
\end{figure}

The fit parameters $\lambda$, $R$ and $\alpha$ are shown in Fig.~\ref{f:0030pars},
as a function of pair $m_T$. We may observe that $\alpha$ is approximately
constant (within systematic uncertainties), and takes an average value of 1.2, being far from the Gaussian
assumption of $\alpha=2$. We also observe, that despite being far from the hydrodynamic limit of Gaussian distributions, the hydro prediction of 
$1/R^2\simeq Am_T+B$ still holds.

The measured correlation function strength $\lambda$ is shown after a normalization by
\begin{align}
\lambda_{\rm max}=\langle \lambda \rangle_{m_{_T}\;=\;0.55-0.9\; {\rm GeV}/c^2}.
\end{align}
This clearly indicates
a decrease at small $m_T$, which may be explained by the larger fraction of low momentum pions coming from
resonance decays. Our measurement is in particular not incompatible with predictions~\cite{Csorgo:2009pa}
based on a reduced $\eta'$ mass, see details in Ref.~\cite{Adare:2017vig}.

We also show, that a new,
empirically found scaling parameter
\begin{align}
\widehat{R}= \frac{R}{\lambda(1+\alpha)}
\end{align}
may be defined. This exhibits a clear
linear scaling with $m_T$, and it also has much decreased statistical uncertainties.

While for the other parameters,
approximate scalings may be observed ($\alpha$ is constant in $m_T$, $1/R^2$ is linear in $m_T$, $\lambda/\lambda_{\rm max}$
follows a unity minus a Gaussian type of $m_T$ dependence), the scaling of $1/\widehat{R}$ is strikingly better than the others.

\section{Centrality and collision energy dependence}

\begin{figure}
\begin{center}
\includegraphics[width=0.495\linewidth]{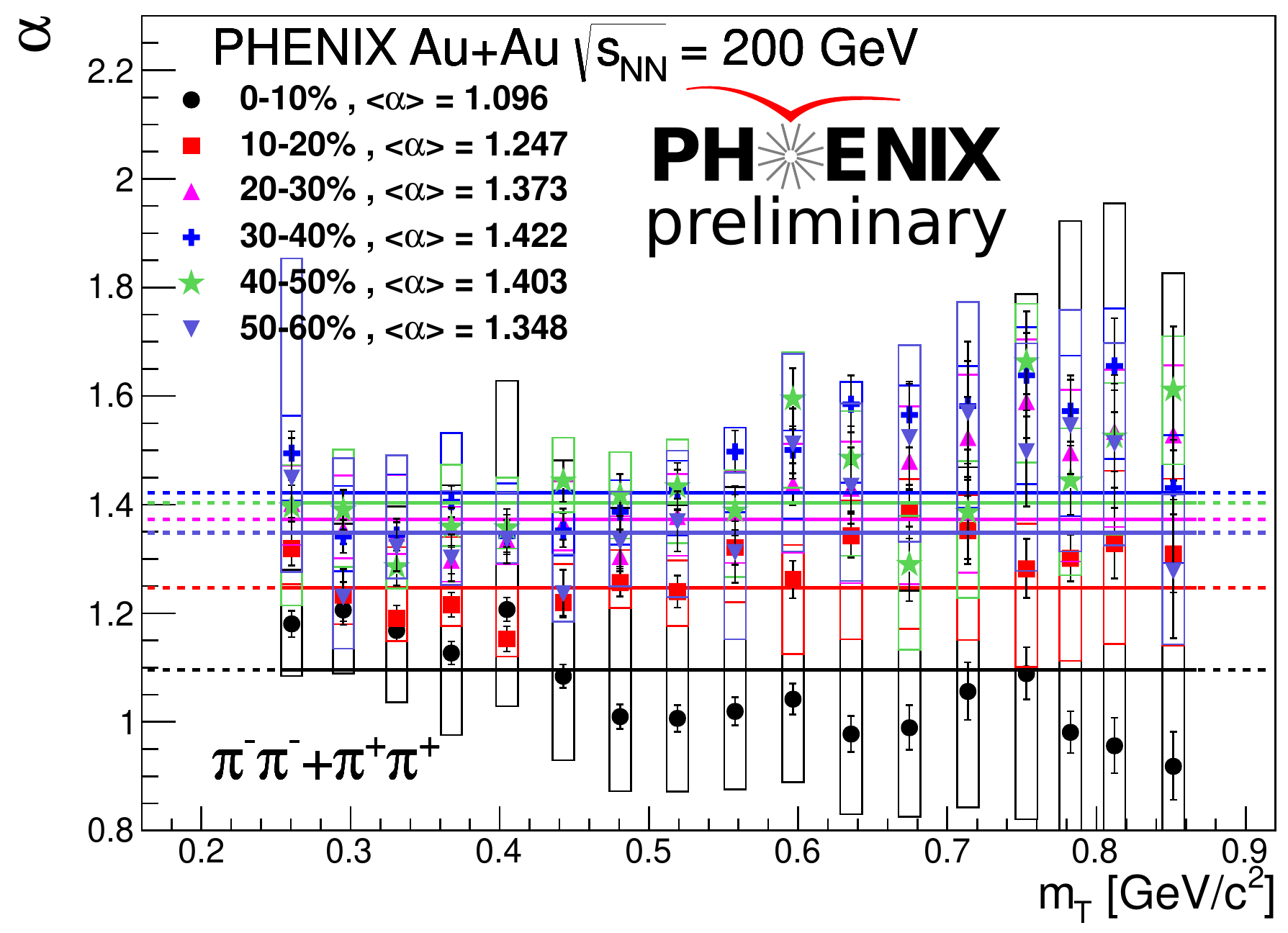}
\includegraphics[width=0.495\linewidth]{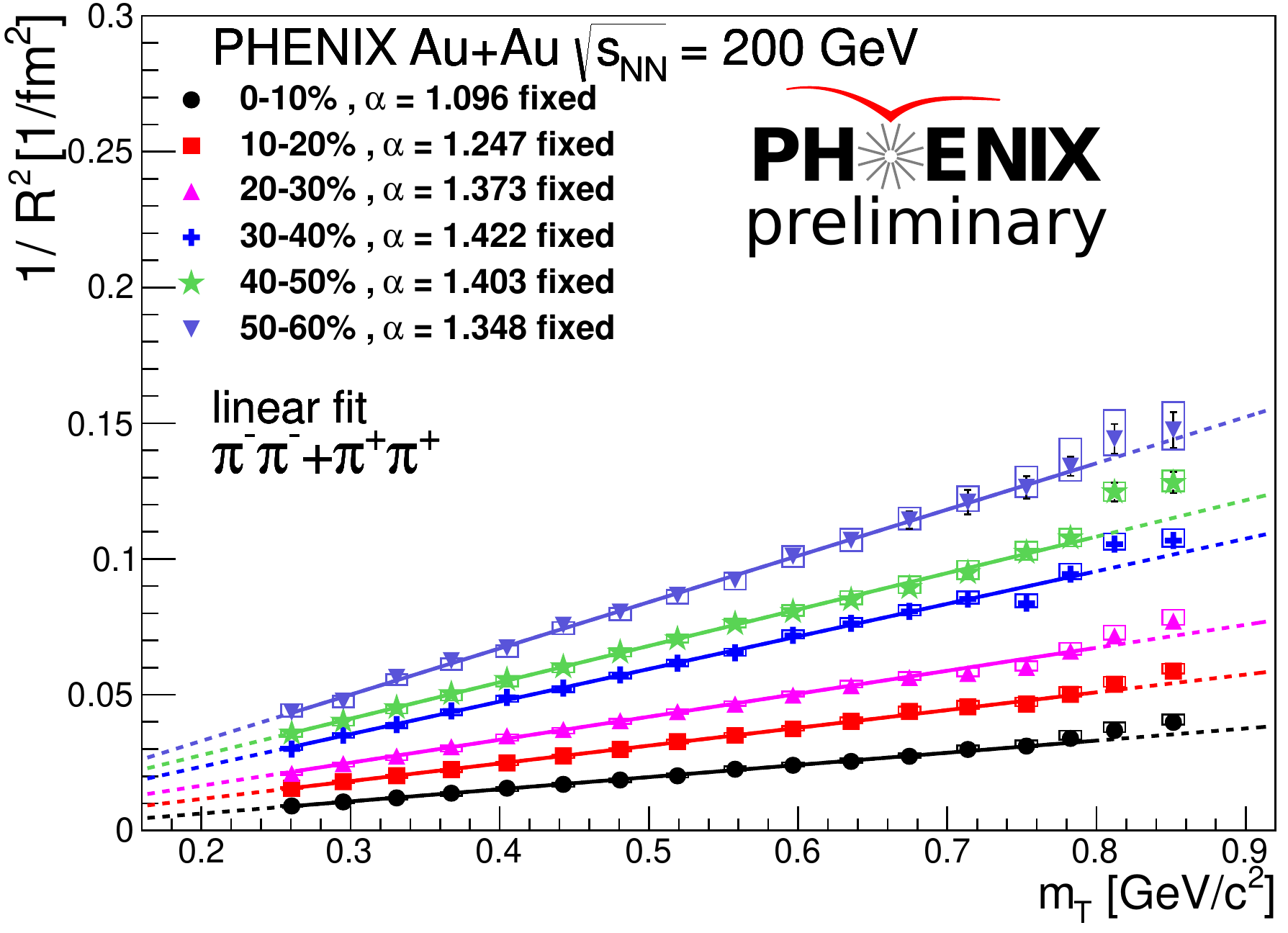}\\
\includegraphics[width=0.495\linewidth]{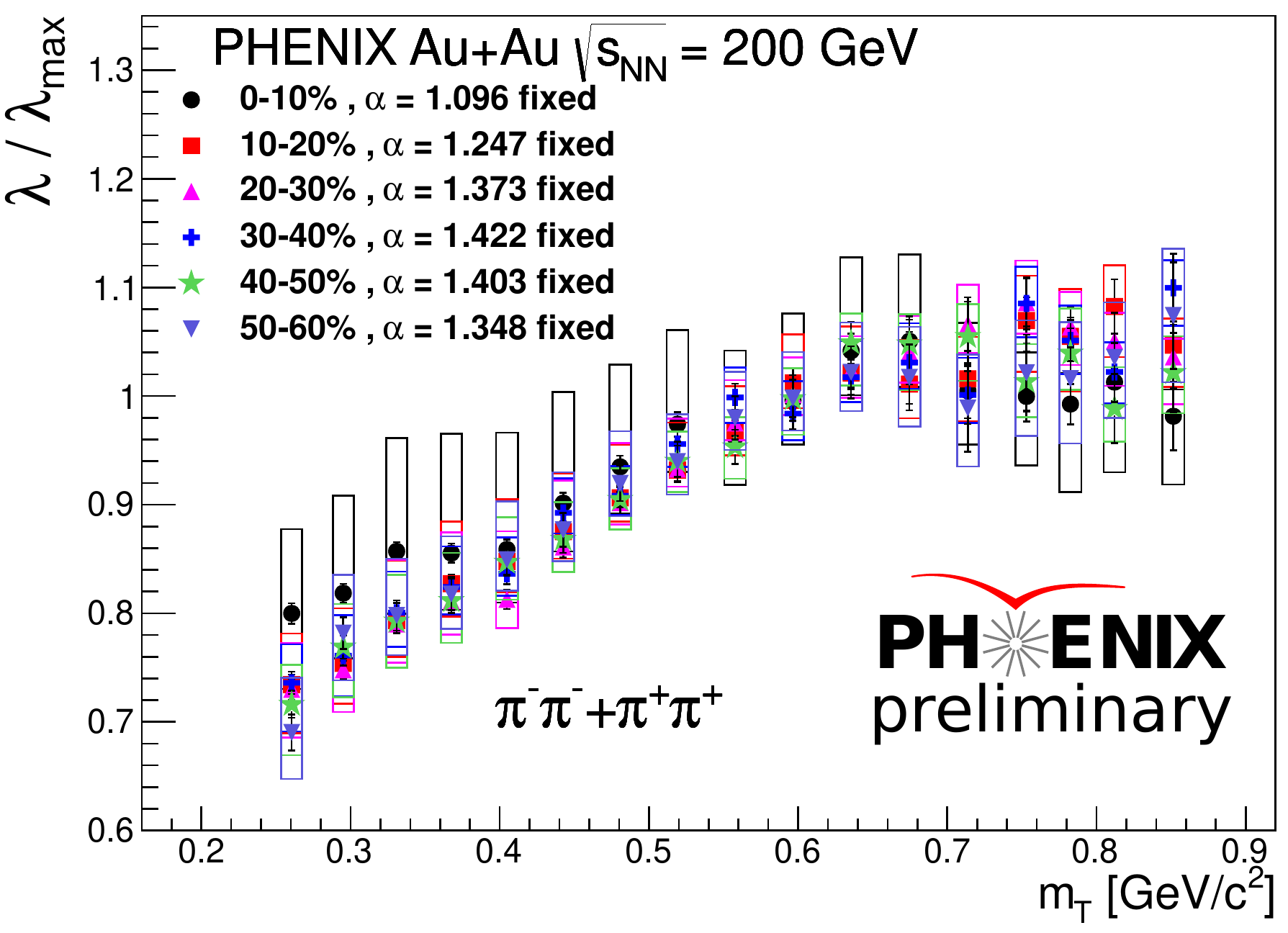}
\includegraphics[width=0.495\linewidth]{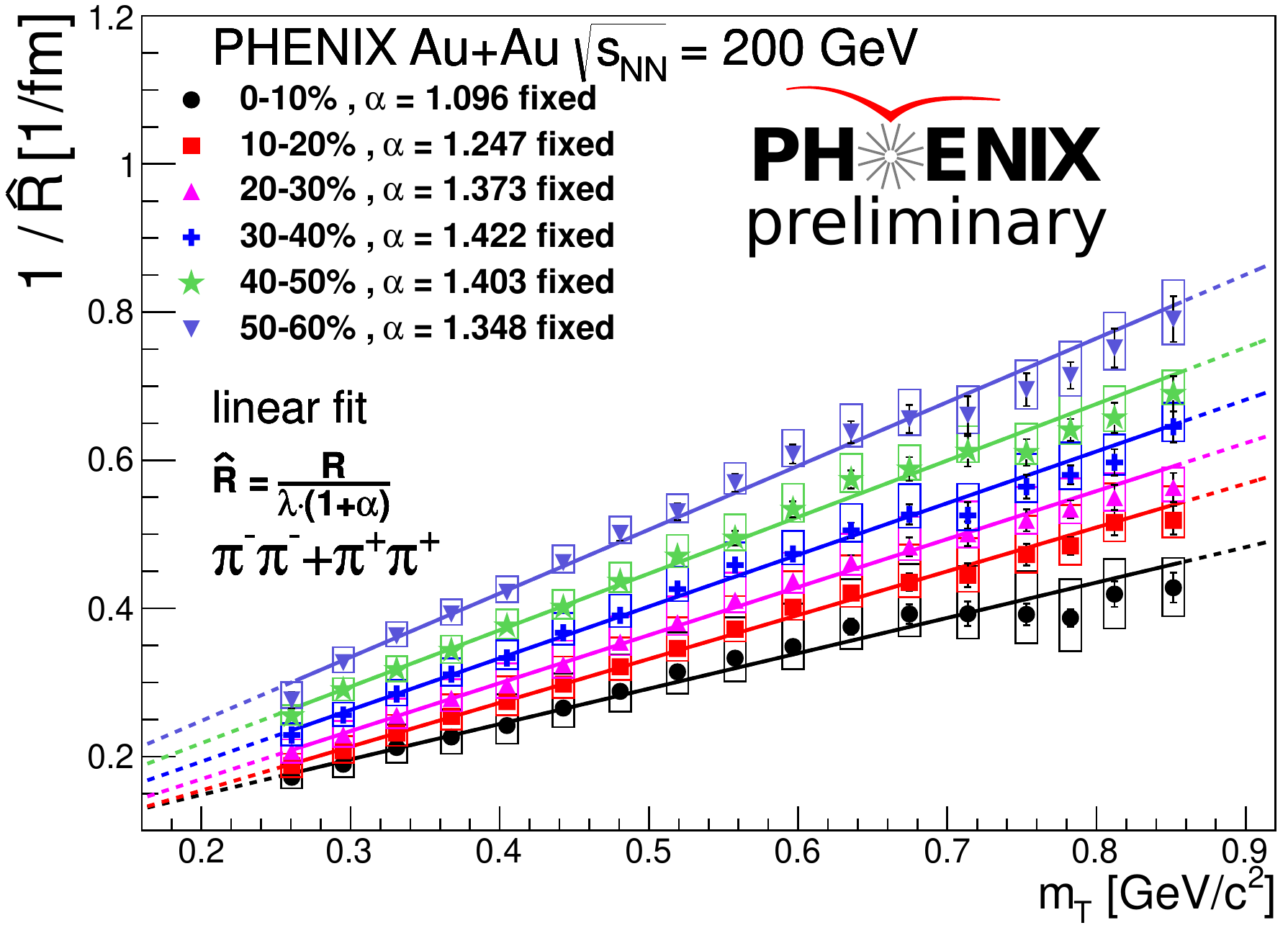}
\end{center}
\caption{Lévy HBT fit parameters for various centralities in 200 GeV Au+Au collisions.\label{f:centdep} }
\end{figure}

\begin{figure}
\begin{center}
\includegraphics[width=0.45\linewidth]{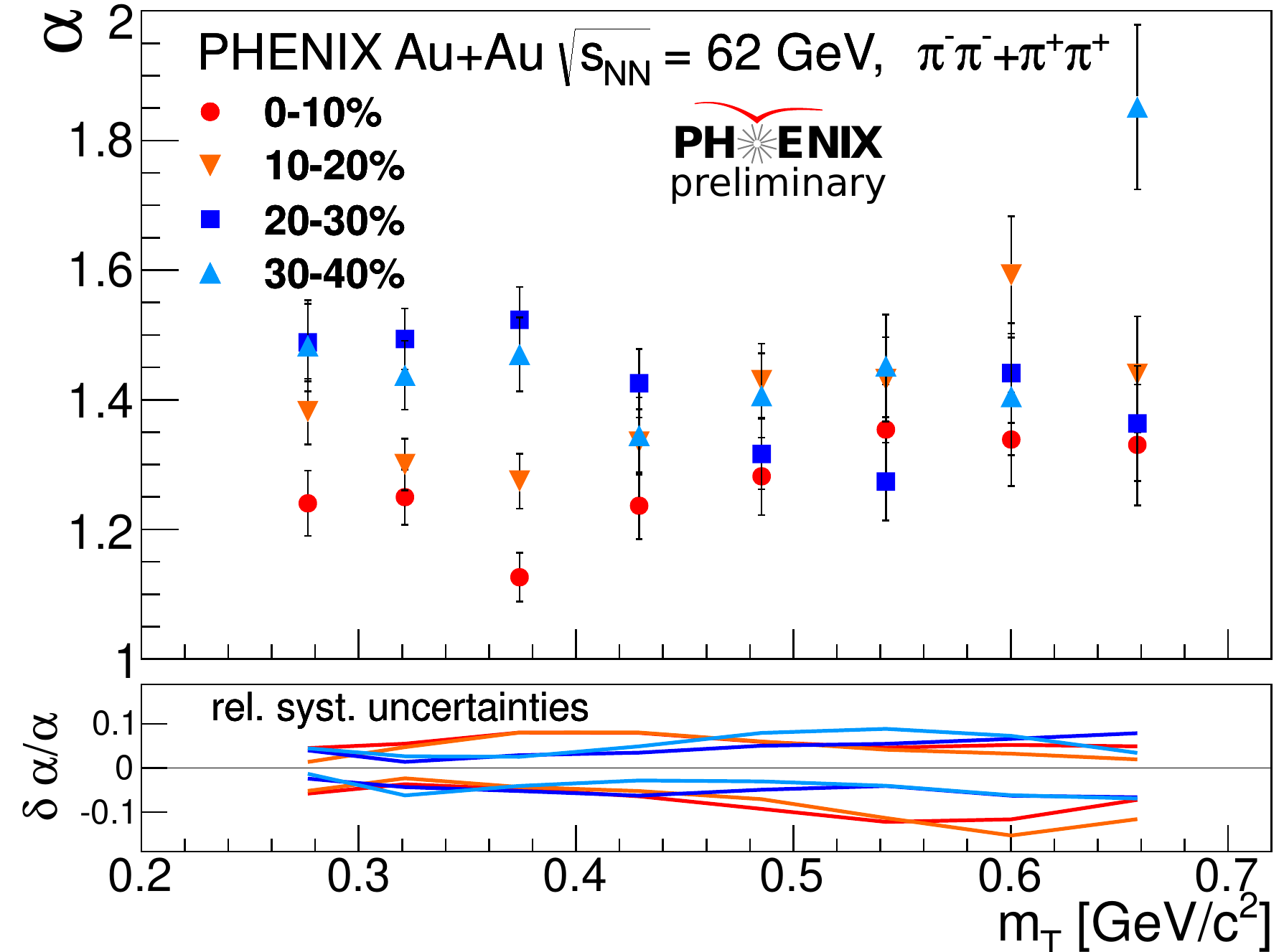}
\includegraphics[width=0.45\linewidth]{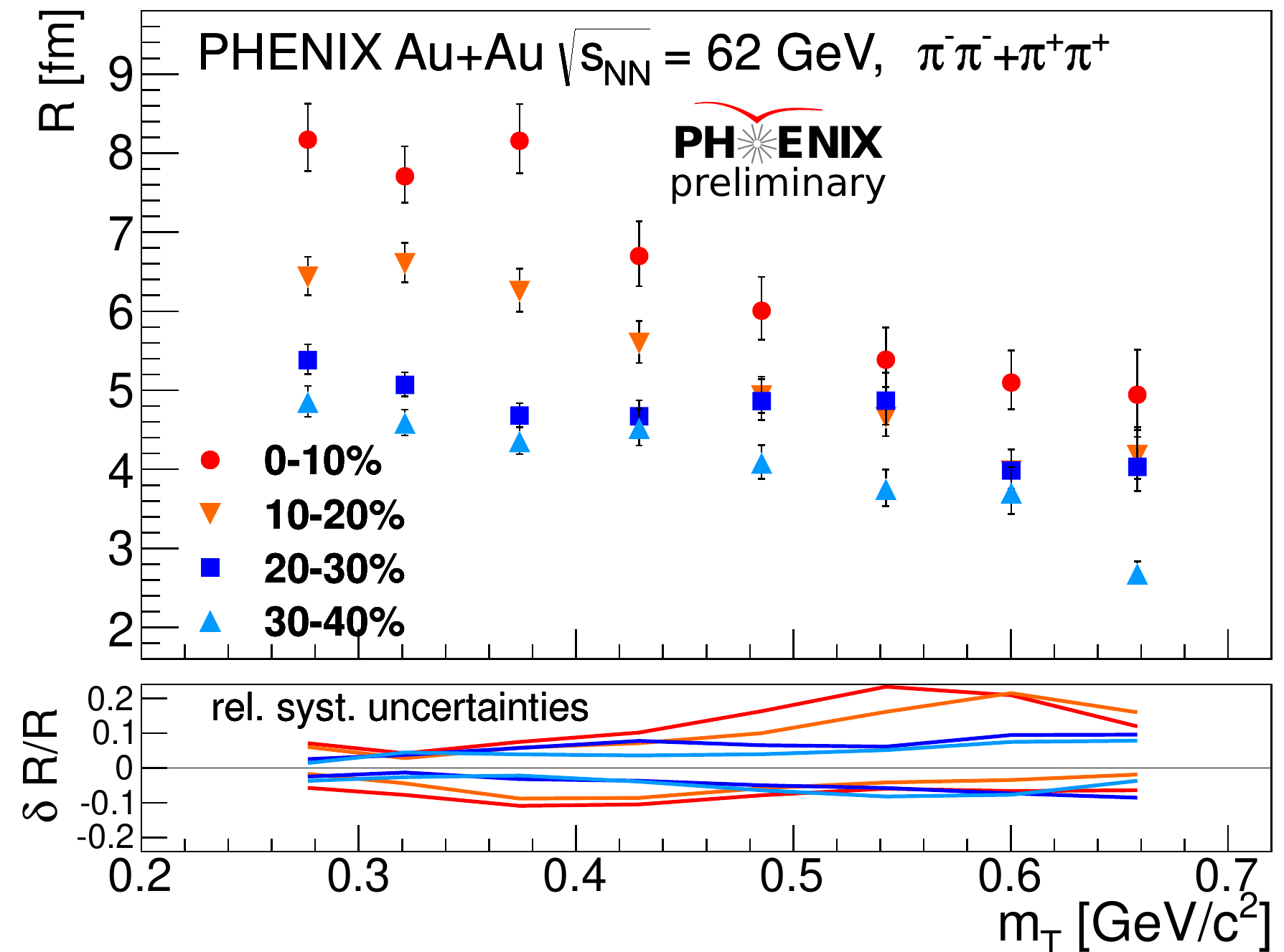}
\includegraphics[width=0.45\linewidth]{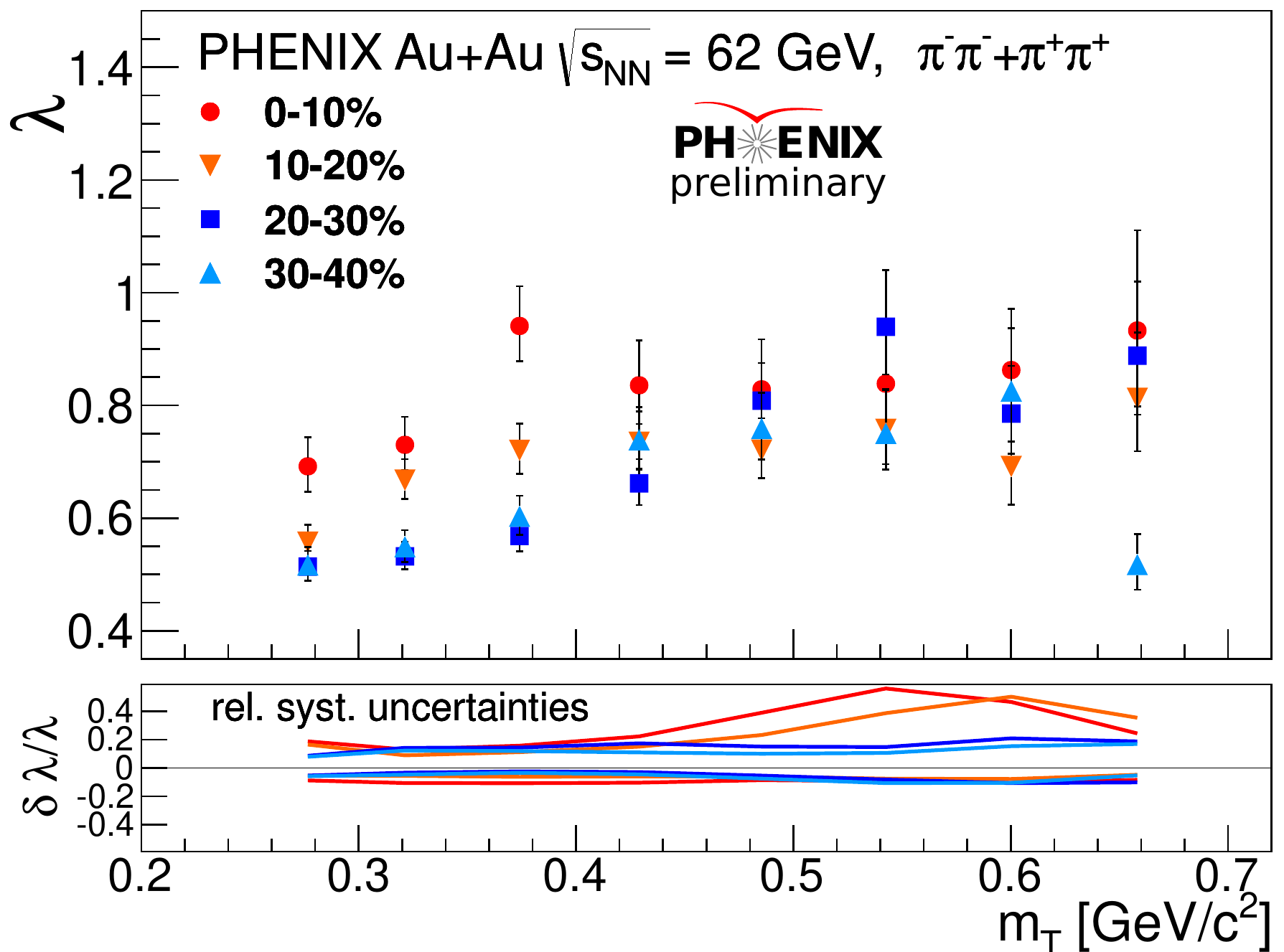}
\includegraphics[width=0.45\linewidth]{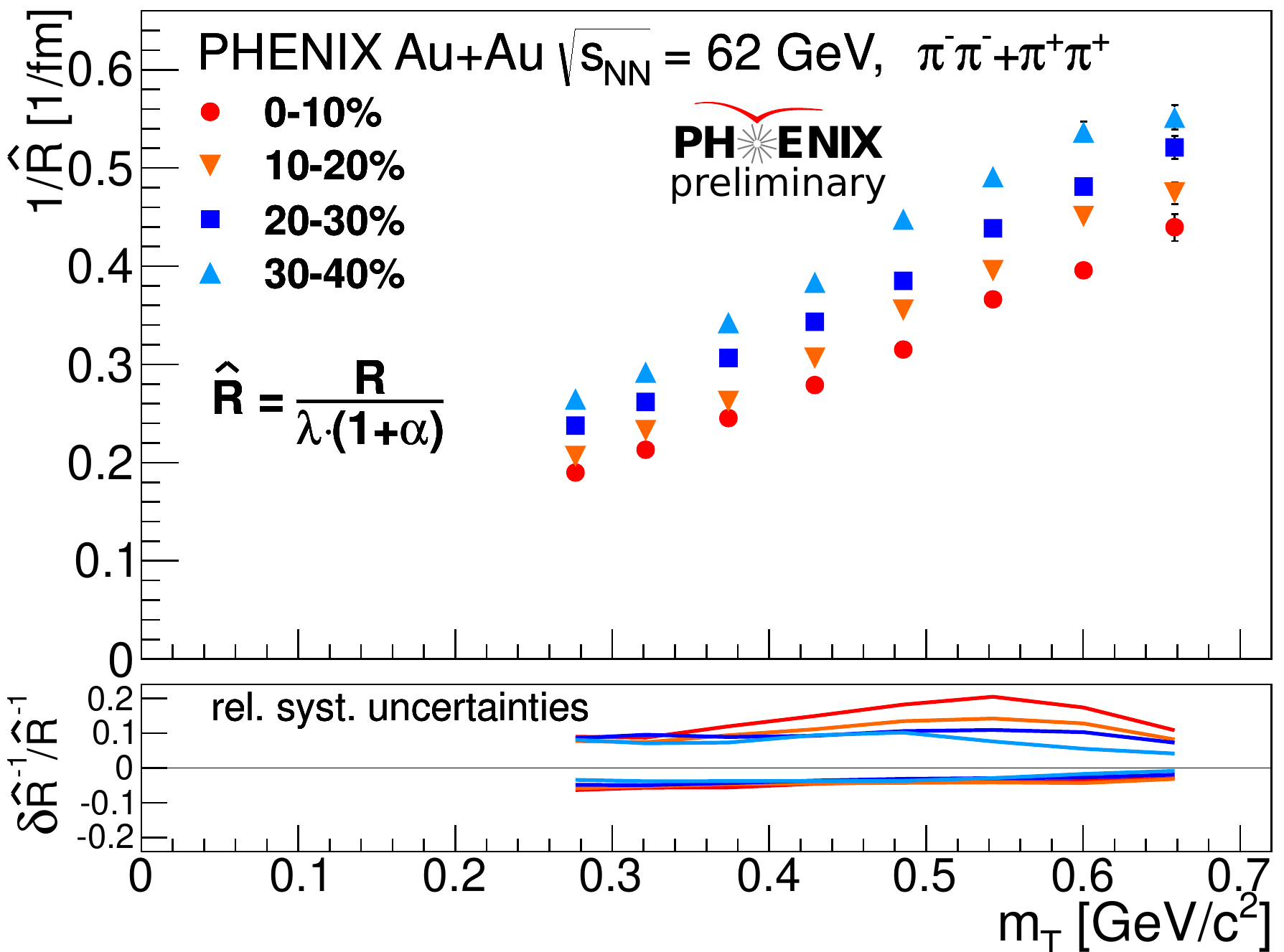}
\includegraphics[width=0.45\linewidth]{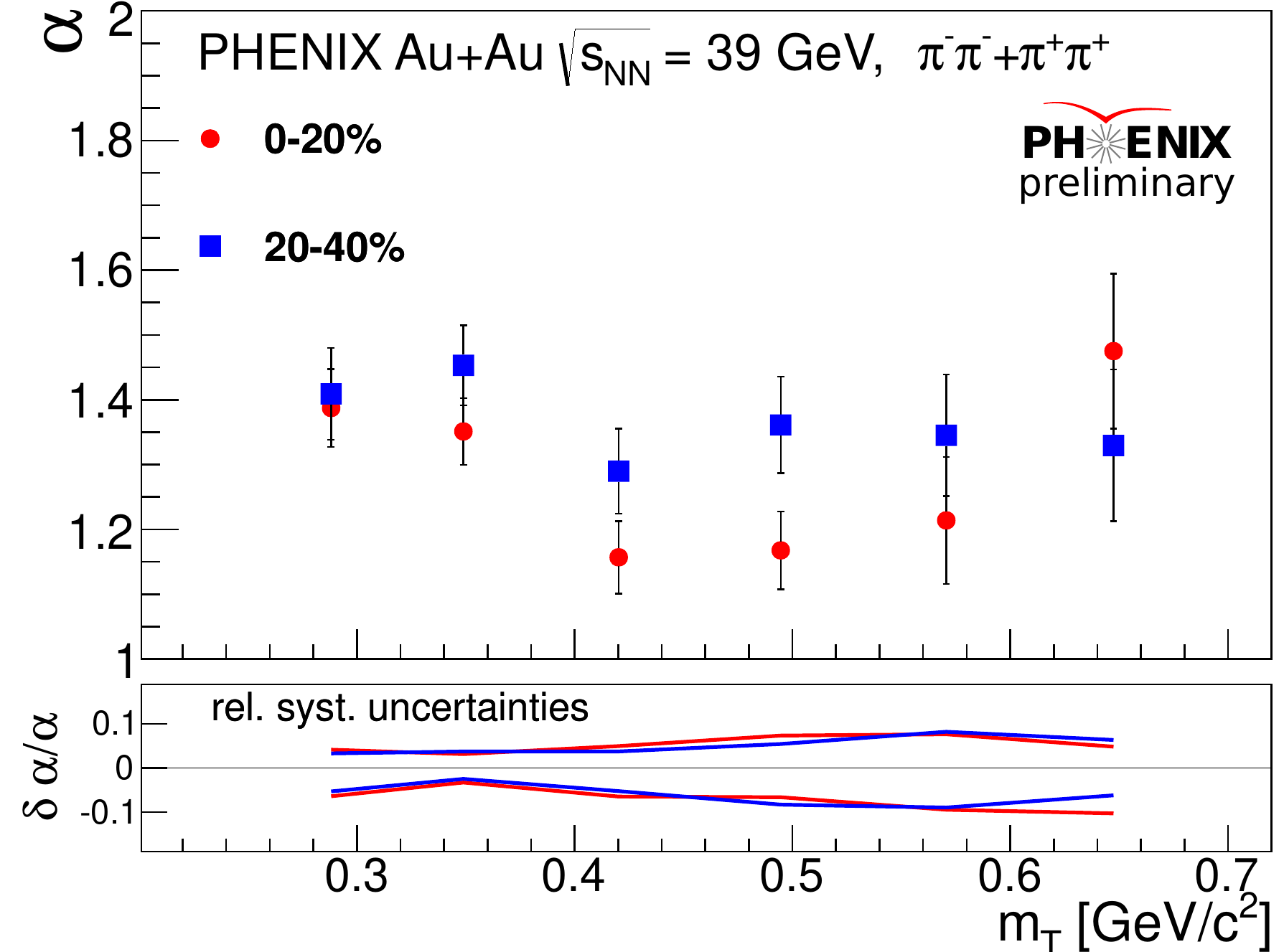}
\includegraphics[width=0.45\linewidth]{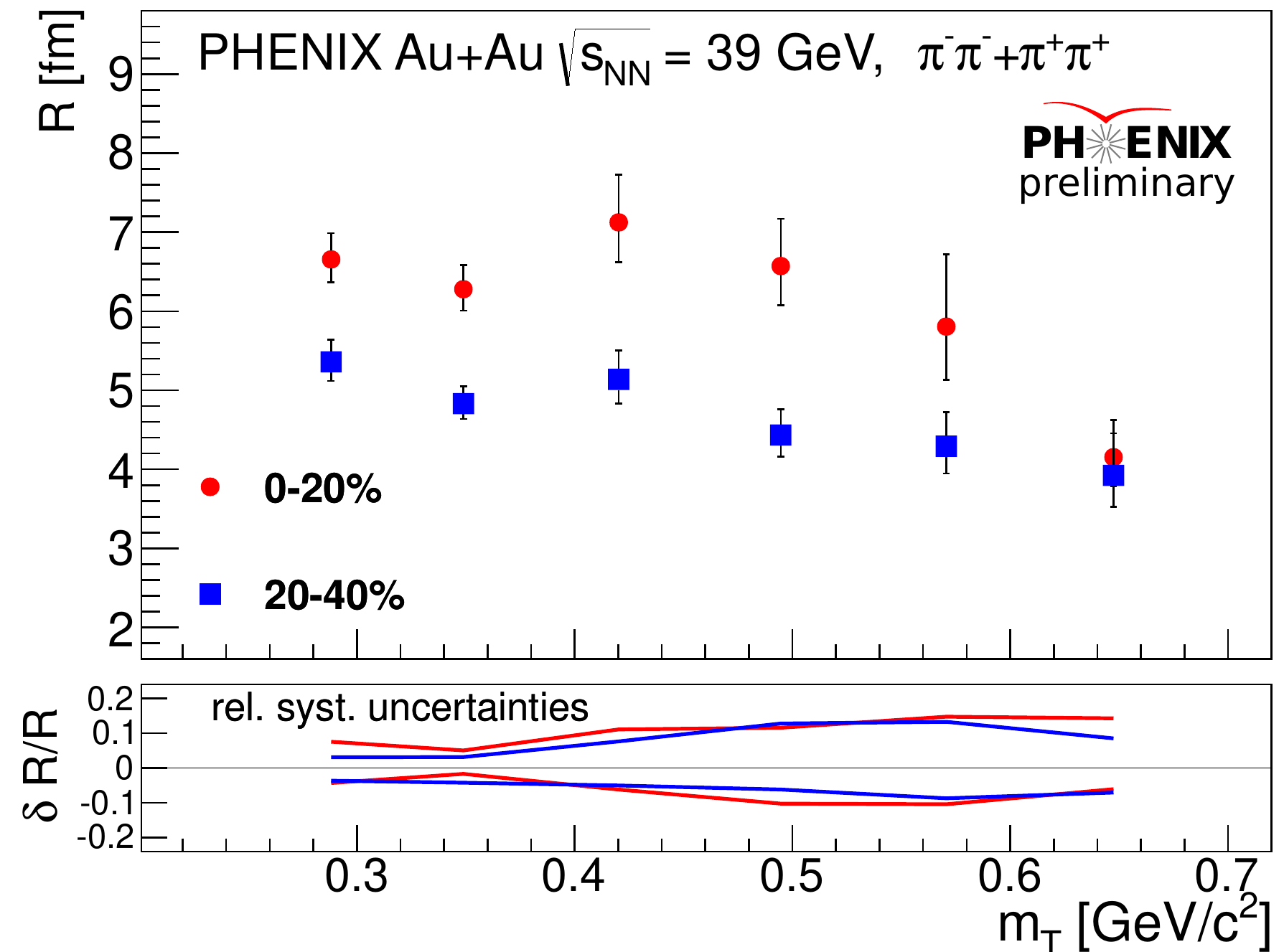}
\includegraphics[width=0.45\linewidth]{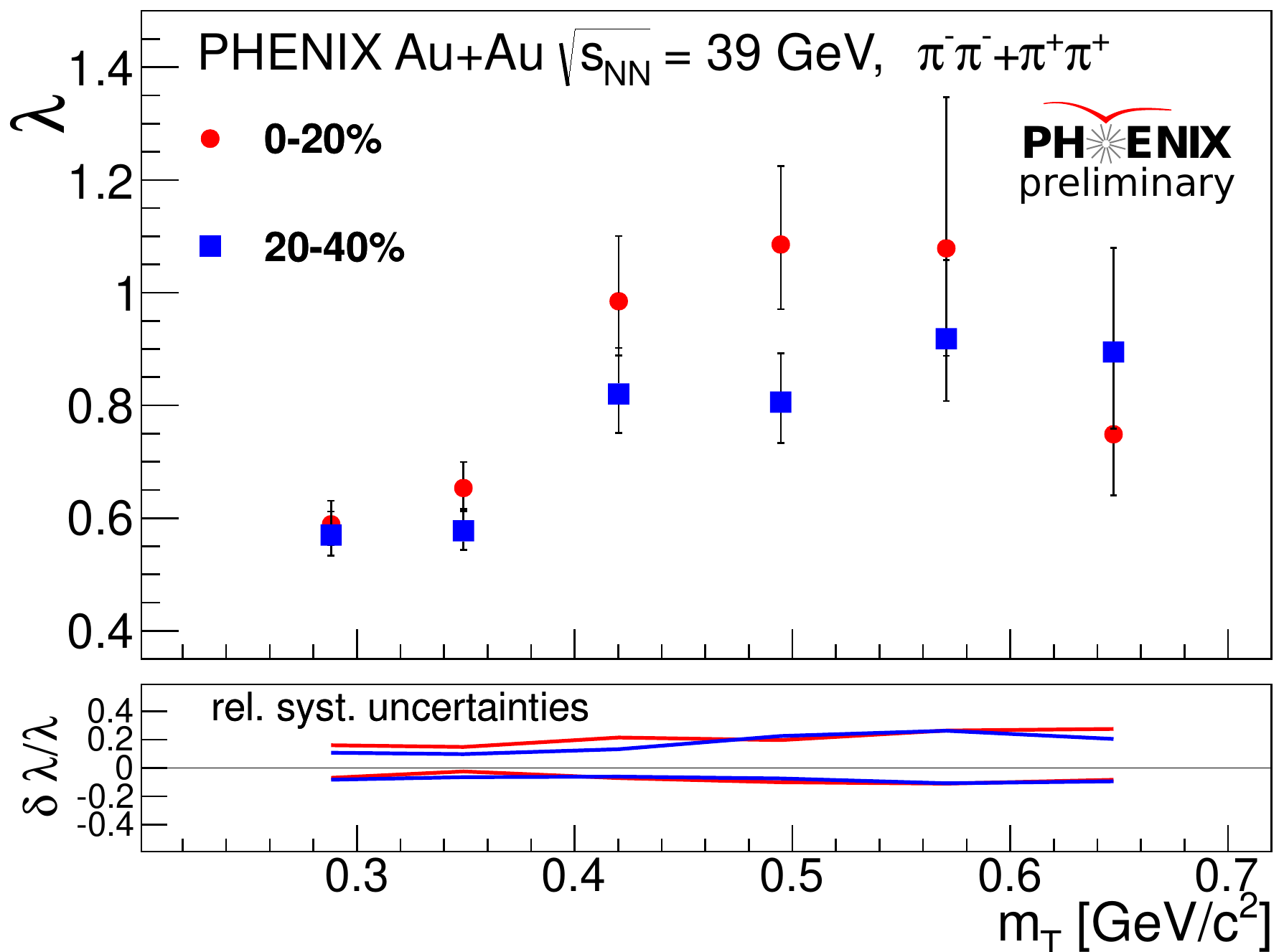}
\includegraphics[width=0.45\linewidth]{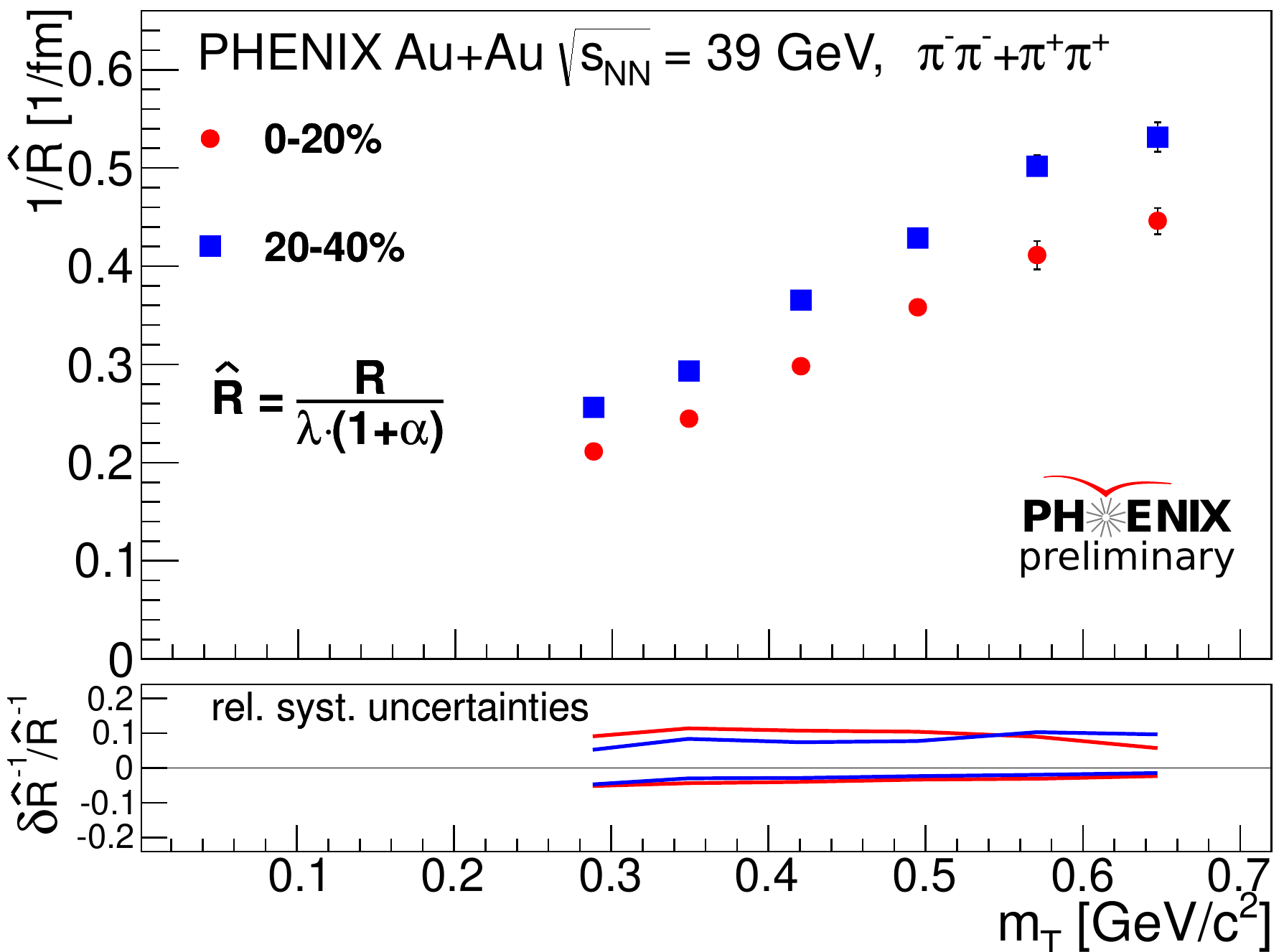}
\end{center}
\caption{Lévy HBT fit parameters in $\sqrt{s_{_{NN}}}=39$ and 62 GeV Au+Au collisions.\label{f:centdep_lowe} }
\end{figure}

We investigated the Lévy HBT correlation functions also at different collision energies, from $\sqrt{s_{_{NN}}}=$ 39 to 200 GeV, and for various
centralities. At 200 GeV, we defined six centrality classes from 0--10\% to 50--60\%, at 62 GeV four classes (from 0--10\% to 30--40\%) and two
classes at 39 GeV (0--20\% and 20--40\%). After performing all the fits to all the measured correlation functions, we extracted the
$m_T$-dependence of the fit parameters. For the 200 GeV analysis, we used 18 $m_T$ bins, 8 bins at 62 GeV and 6 bins at 39 GeV.
The extracted fit parameters (as a function of $m_T$) are shown in Figs.~\ref{f:centdep}-\ref{f:centdep_lowe}. At each energy, we observe 
an approximately $m_T$ independent $\alpha$ value. To investigate the $m_T$ dependence of the parameters, we performed fits with $\alpha$
fixed to its average value in the given centrality bin. This resulted in the clear trends shown in Fig.~\ref{f:centdep}: in each centrality,
$1/R^2$ and $1/\widehat{R}$ scale linearly with $m_T$, and a ``hole'' of similar magnitude is observed for $\lambda/\lambda_{\rm max}$.
At the lower energies (shown in Fig.~\ref{f:centdep_lowe}), we observe qualitatively similar trends, but statistics prevents us
from drawing more definitive conclusions. In both case, the most clear trend is observable for the scaling variable $\widehat{R}$, which is
apparently the less prone to statistical fluctuations of the correlation function and the fit.

\section{Three-pion correlations}

\begin{figure}
\centering
\includegraphics[width=0.67\linewidth]{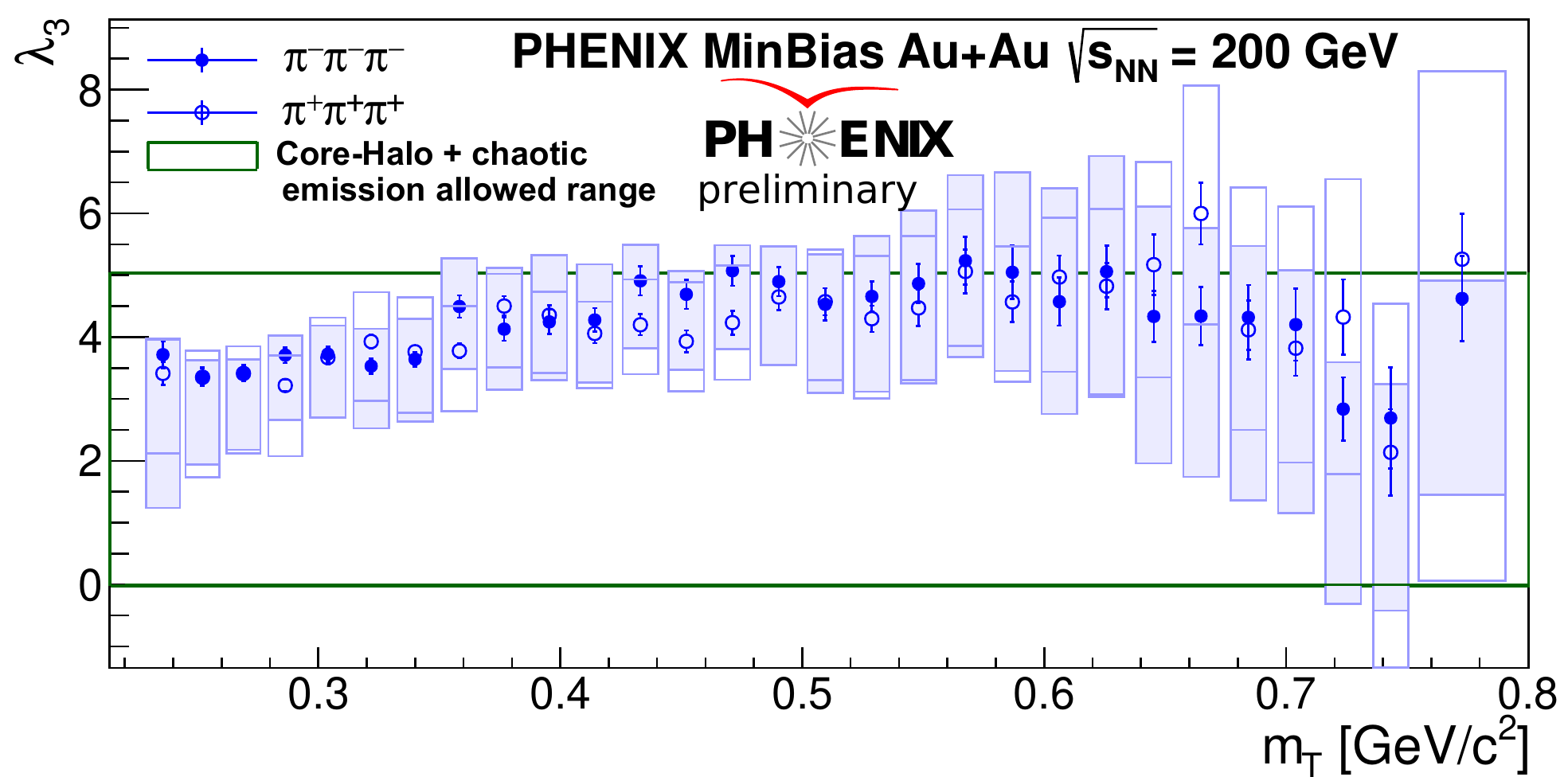}
\includegraphics[width=0.7\linewidth]{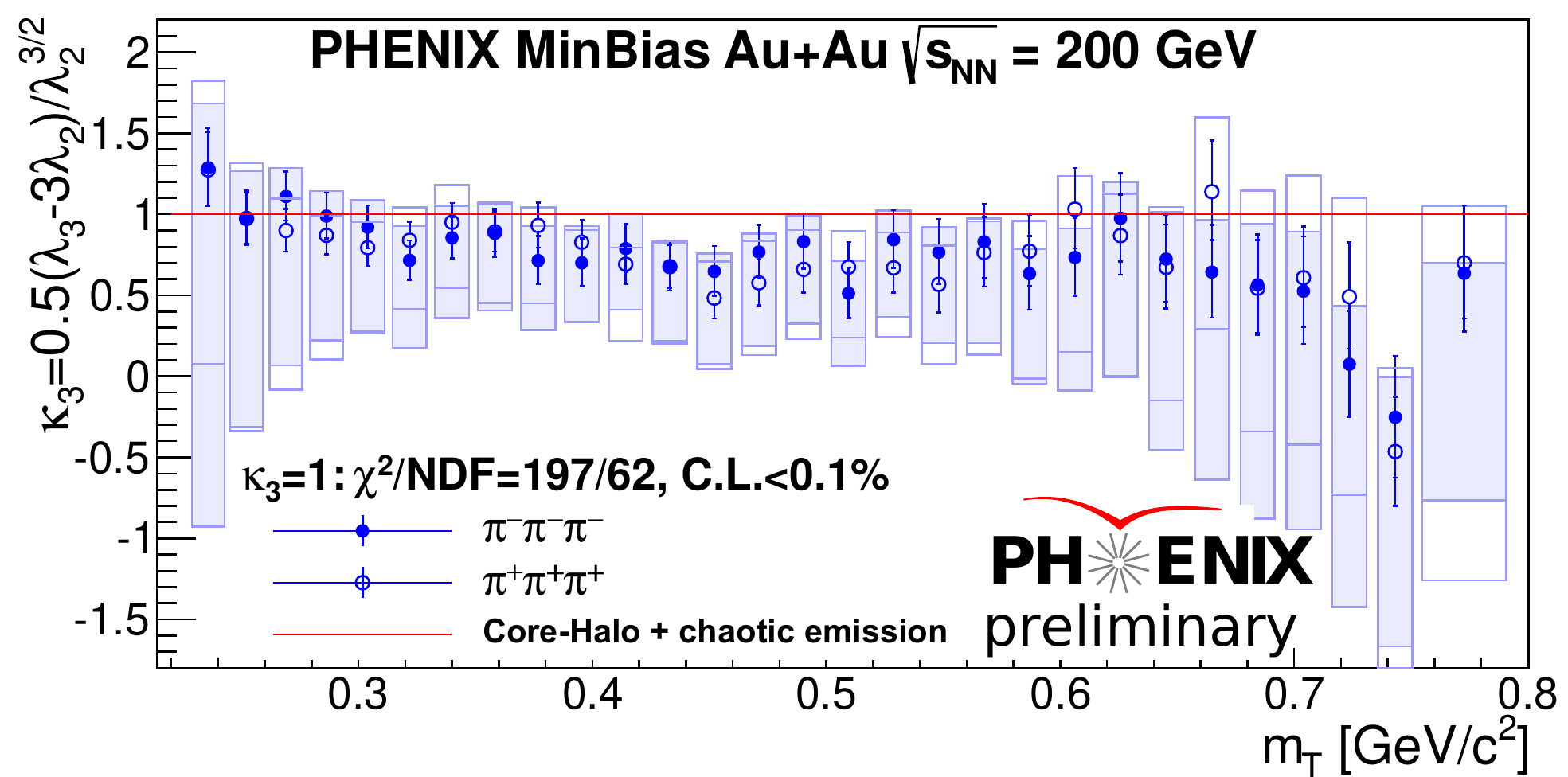}
\caption{The three-particle correlation strength, $\lambda_3$, as function of $m_T$ (top panel).
Core-halo independent parameter $\kappa_3$ as function of $m_T$ (bottom panel).\label{f:kappa3}}
\end{figure}

As mentioned above, the $\lambda$ intercept parameter of the two-pion momentum correlations
may be related to the fraction of pions coming from the decays of long lived resonances. If so,
the possible enhancement of $\eta'$ production may be connected to it. An important cross-check
of this can be performed by the measurement of multi-pion correlations, since their zero momentum 
intercept is also related to the core/halo ratio. It turns out~\cite{Csorgo:1999sj} that if we define
two- and three-particle correlation intercepts (neglecting $K$ dependence) as
\begin{align}
\lambda_2 &\equiv C_2(Q_{12}\rightarrow 0)-1\\
\lambda_3 &\equiv C_3(Q_{12}\rightarrow 0,Q_{13}\rightarrow 0,Q_{23}\rightarrow 0)-1
\end{align}  
then these can be expressed from the core fraction $f_c$ as
\begin{align}
\lambda_2= f_c^2, \qquad \lambda_3= 2f_c^3+3f_c^2.
\end{align}
It is immediately clear from this, that we may define 
\begin{align}
\kappa_3=\frac{\lambda_3 - 3\lambda_2}{2\sqrt{\lambda_2}^3}
\end{align}
which is equal to one in case of the core-halo model. Hence if $\kappa_3\neq 1$, that would point to
pion production beyond the core-halo model, such as coherence, as also discussed in Ref.~\cite{Novak:2018hqd}.
We measured the value of the above defined $\lambda_3$ and $\kappa_3$ parameters, as shown in Fig.~\ref{f:kappa3}.
The preliminary uncertainties were calculated from simultaneous variation of two- and three-pion analysis
settings. We observe that $\kappa_3$ is near unity, but the preliminary status of this analysis prevents us 
from making a more definitive conclusion.

\section{Summary}
In summary, we measured two- and three-particle correlation functions of identical pions at various
collision energies and centralities, and fitted them with correlation functions calculated from Lévy
sources. We found that this assumption yields a statistically acceptable description of all results.
We extracted the $m_T$ dependence of two-particle Lévy parameters $\alpha$, $R$ and $\lambda$.
We found that the $\alpha$ parameter is far from 2, and has only a slight $m_T$ dependence, but
a more pronounced centrality dependence. We also found that $1/R^2$ is approximately linear
with $m_T$, while $\lambda$ decreases at low $m_T$, showing a ``hole-like'' structure, not incompatible
with models utilizing a decreased $\eta'$ mass. We also demonstrated the appearance of a scaling
parameter $\widehat{R}$, the inverse of which shows a very  clear linear scaling with $m_T$. Furthermore,
we investigated if the interpretation of $\lambda$ holds as being related to the core-halo ratio, by measuring
three-particle correlations. The $\kappa_3$ parameter, which equals to unity in the core-halo model, was 
measured to be close to this conjectured value.

\section*{Acknowledgments}
The author expresses gratitude for the excellent organization of the WWND 2018 conference.
The author was supported by the Hungarian NKIFH grant No. FK-123842,
by the János Bolyai Research Scholarship of the Hungarian Academy of Sciences and the ÚNKP-17-4 New National Excellence
Program of the Hungarian Ministry of Human Capacities. 

\section*{References}
\bibliographystyle{iopart-num}
\bibliography{../../../Master}

\providecommand{\newblock}{}
\begin{thebibliography}{10}
\expandafter\ifx\csname url\endcsname\relax
  \def\url#1{{\tt #1}}\fi
\expandafter\ifx\csname urlprefix\endcsname\relax\def\urlprefix{URL }\fi
\providecommand{\eprint}[2][]{\url{#2}}

\bibitem{HanburyBrown:1952na}
Hanbury~Brown R, Jennison R~C and K D~G~M 1952 {\em Nature\/} {\bf 170}
  1061–1063

\bibitem{HanburyBrown:1956zza}
Brown R~H and Twiss R 1956 {\em Nature\/} {\bf 177} 27--29

\bibitem{HanburyBrown:1957na}
Twiss R~Q, Little A~G and Hanbury~Brown R 1957 {\em Nature\/} {\bf 180} pages
  324–326

\bibitem{HanburyBrown:1956pf}
Hanbury~Brown R and Twiss R~Q 1956 {\em Nature\/} {\bf 178} 1046--1048

\bibitem{Townes:1999apj}
Townes C~H 1999 {\em Astrophysical Journal\/} {\bf 525C} 148

\bibitem{Michelson:1921apj}
Michelson A~A and Pease F~G 1921 {\em Astrophysical Journal\/} {\bf 53}
  249--259

\bibitem{HanburyBrown:1967mna}
Brown R~H, Davis J and Allen L~R 1967 {\em Mon. Notices Royal Astron. Soc.\/}
  {\bf 137} 375--392

\bibitem{HanburyBrown:1964na}
Hanbury~Brown R, Hazard C, Davis J and Allen L~R 1964 {\em Nature\/} {\bf 201}
  pages 1111--1112

\bibitem{HanburyBrown:1967mnb}
Brown R~H, Davis J, Allen L~R and Rome J~M 1967 {\em Mon. Notices Royal Astron.
  Soc.\/} {\bf 137} 393--417

\bibitem{Dravins:2010spi}
Dravins D, Jensen H, LeBohec S and Nu{\~n}ez P~D 2010 {\em Optical and Infrared
  Interferometry II\/} ({\em Proceedings of the SPIE\/} vol 7734) p 77340A
  (\textit{Preprint} \eprint{1009.5815})

\bibitem{Glauber:1962tt}
Glauber R~J 1963 {\em Phys. Rev. Lett.\/} {\bf 10} 84--86

\bibitem{Glauber:2006zz}
Glauber R~J 2006 {\em Rev. Mod. Phys.\/} {\bf 78} 1267--1278

\bibitem{Glauber:2006gd}
Glauber R~J 2006 {\em Nucl. Phys.\/} {\bf A774} 3--13 (\textit{Preprint}
  \eprint{nucl-th/0604021})

\bibitem{Goldhaber:1959mj}
Goldhaber G, Fowler W~B, Goldhaber S and Hoang T~F 1959 {\em Phys. Rev.
  Lett.\/} {\bf 3} 181--183

\bibitem{Goldhaber:1960sf}
Goldhaber G, Goldhaber S, Lee W~Y and Pais A 1960 {\em Phys. Rev.\/} {\bf 120}
  300--312

\bibitem{Lednicky:2001qv}
Lednicky R 2001  (\textit{Preprint} \eprint{nucl-th/0112011})

\bibitem{Adare:2017vig}
Adare A {\em et~al.\/} (PHENIX) 2017  (\textit{Preprint} \eprint{1709.05649})

\bibitem{Metzler:1999zz}
Metzler R, Barkai E and Klafter J 1999 {\em Phys. Rev. Lett.\/} {\bf 82}
  3563--3567

\bibitem{Csanad:2007fr}
Csan\'ad M, Cs\"org\H{o} T and Nagy M 2007 {\em Braz. J. Phys.\/} {\bf 37}
  1002--1013 (\textit{Preprint} \eprint{hep-ph/0702032})

\bibitem{Csorgo:2003uv}
Cs\"org\H{o} T, Hegyi S and Zajc W~A 2004 {\em Eur. Phys. J.\/} {\bf C36}
  67--78 (\textit{Preprint} \eprint{nucl-th/0310042})

\bibitem{Csanad:2017usp}
Csan{\'a}d M (PHENIX) 2017 {\em Universe\/} {\bf 3} 85 (\textit{Preprint}
  \eprint{1711.05575})

\bibitem{Kincses:2017zlb}
Kincses D (PHENIX) 2018 {\em Universe\/} {\bf 4} 11 (\textit{Preprint}
  \eprint{1711.06891})

\bibitem{Lokos:2018dqq}
L{\"o}k{\"o}s S (PHENIX) 2018 {\em Universe\/} {\bf 4} 31 (\textit{Preprint}
  \eprint{1801.08827})

\bibitem{Novak:2018hqd}
Nov{\'a}k T (PHENIX) 2018 {\em Universe\/} {\bf 4} 57 (\textit{Preprint}
  \eprint{1801.03544})

\bibitem{Bolz:1992hc}
Bolz J, Ornik U, Plumer M, Schlei B and Weiner R 1993 {\em Phys.Rev.\/} {\bf
  D47} 3860--3870

\bibitem{Csorgo:1994in}
Cs\"org\H{o} T, L\"orstad B and Zim\'anyi J 1996 {\em Z. Phys.\/} {\bf C71}
  491--497 (\textit{Preprint} \eprint{hep-ph/9411307})

\bibitem{Kapusta:1995ww}
Kapusta J~I, Kharzeev D and McLerran L~D 1996 {\em Phys. Rev.\/} {\bf D53}
  5028--5033 (\textit{Preprint} \eprint{hep-ph/9507343})

\bibitem{Vance:1998wd}
Vance S~E, Cs\"org\H{o} T and Kharzeev D 1998 {\em Phys. Rev. Lett.\/} {\bf 81}
  2205--2208 (\textit{Preprint} \eprint{nucl-th/9802074})

\bibitem{Csorgo:2009pa}
Cs\"org\H{o} T, V\'ertesi R and Sziklai J 2010 {\em Phys.Rev.Lett.\/} {\bf 105}
  182301 (\textit{Preprint} \eprint{0912.5526})

\bibitem{Adcox:2004mh}
Adcox K {\em et~al.\/} (PHENIX Collaboration) 2005 {\em Nucl. Phys.\/} {\bf
  A757} 184--283 (\textit{Preprint} \eprint{nucl-ex/0410003})

\bibitem{Csorgo:1999sj}
Cs\"org\H{o} T 2002 {\em Heavy Ion Phys.\/} {\bf 15} 1--80 (\textit{Preprint}
  \eprint{hep-ph/0001233})

\end{thebibliography}

\end{document}